\documentclass[conference]{IEEEtran}

\usepackage{cite}            % For managing multiple citations
\usepackage{amsmath,amssymb,amsfonts}  % For mathematical symbols
\usepackage{algorithmic}     % For algorithms
\usepackage{graphicx}        % For including figures
\usepackage{textcomp}        % Additional text symbols
\usepackage{xcolor}          % For colored text
\usepackage{lipsum}          % Generates dummy text
\usepackage{multirow}
\usepackage{array}
\usepackage{subcaption}
\usepackage{float}
\usepackage{caption}
\usepackage{tabularx}
\usepackage{booktabs}
\usepackage{multirow}
\usepackage{multicol}

\begin{document}

\title{Comprehensive mm-Wave FMCW Radar Dataset for Vital Sign Monitoring: Embracing Extreme Physiological Scenarios}

\author{\IEEEauthorblockN{Ehsan Sadeghi}
\IEEEauthorblockA{\textit{EEMCS Faculty} \\
\textit{University of Twente}\\
Enschede, The Netherlands \\
e.sadeghi@utwente.nl}
\and
\IEEEauthorblockN{Karina Skurule}
\IEEEauthorblockA{\textit{EEMCS Faculty} \\
\textit{University of Twente}\\
Enschede, The Netherlands \\
k.m.skurule@student.utwente.nl}
\and
\IEEEauthorblockN{Alessandro Chiumento}
\IEEEauthorblockA{\textit{EEMCS Faculty} \\
\textit{University of Twente}\\
Enschede, The Netherlands \\
a.chiumento@utwente.nl}
\and
\IEEEauthorblockN{Paul Havinga}
\IEEEauthorblockA{\textit{EEMCS Faculty} \\
\textit{University of Twente}\\
Enschede, The Netherlands \\
p.j.m.havinga@utwente.nl}
}

\maketitle

\begin{abstract}
Recent advancements in non-invasive health monitoring technologies underscore the potential of mm-Wave Frequency-Modulated Continuous Wave (FMCW) radar in real-time vital sign detection. This paper introduces a novel dataset, the first of its kind, derived from mm-Wave FMCW radar, meticulously capturing heart rate and respiratory rate under various conditions. Comprising data from ten participants, including scenarios with elevated heart rates and participants with diverse physiological profiles such as asthma and meditation practitioners, this dataset is validated against the Polar H10 sensor, ensuring its reliability for scientific research. This dataset can offer a significant resource for developing and testing algorithms aimed at non-invasive health monitoring, promising to facilitate advancements in remote health monitoring technologies.
\end{abstract}


\section{Introduction}
\IEEEPARstart{M}{onitoring} vital signs, including heart rate (HR) and respiratory rate (RR), plays a pivotal role in the management and prevention of numerous health conditions. Traditionally, such monitoring has been carried out within hospital settings, with established protocols reflecting the importance of these parameters in assessing patient health \cite{lockwood2004vital}. However, in some cases patient vital sign monitoring at home is required \cite{mok2015vital}. Wearable monitoring systems are widely used because of their low costs and accuracy. There has been a significant increase in the development and use of wearable devices for monitoring vital signs in humans. Wearable devices such as smartwatches (like the VitalTracer and the Fitbit) and chest straps can continuously monitor vital signs and provide real-time feedback to the user. Non-wearable devices such as camera-based systems \cite{antink2019broader}, radar systems \cite{li2008random}, and hybrid systems \cite{gu2013hybrid} have also been developed to monitor vital signs in humans in a non-invasive way.

Radar technology has emerged as a promising tool for a wide array of applications in both human and animal health monitoring, particularly excelling in vital sign estimation \cite{RadarP, new-AR, new-vsign,raypet}.
% The advantages are the non-necessity of a direct skin contact, and to remove clothes, and radars are less affected by environmental conditions like light, temperature, and humidity when compared to other non-wearable sensors such as camera and infrared sensors.
The benefits include the elimination of the need for direct skin contact and disrobing, along with radars' reduced sensitivity to environmental factors such as light, temperature, and humidity, in contrast to other non-wearable sensors like cameras and infrared sensors.
Furthermore, radar-based sensing addresses existing privacy concerns and issues present with other non-invasive technologies, such as cameras \cite{RadarP}.
In literature, the utilization of radar technology for vital sign detection has demonstrated promising outcomes and holds the potential to enhance the management of various medical conditions. To this end, researchers have employed different types of radar systems to investigate challenges associated with vital sign detection. For instance, ultra wide band radar operating at 24 GHz \cite{sakamoto2015feature}, stepped frequency continuous wave radar spanning 2-4 GHz \cite{ren2015non}, continuous wave radar at 14 GHz \cite{lee201314}, and millimeter Wave (mm-wave) Frequency Modulated Continuous Wave (FMCW) radar operating between 75-85 GHz \cite{wang2015novel} have been explored.

Within the spectrum of radar technologies, FMCW radar has gained significant attention for its proficiency in determining the distance, angle, and speed of objects. Specifically, mm-wave FMCW radar, operating within the higher frequency ranges of 30 GHz to 300 GHz (corresponding to wavelengths of 1 mm to 10 mm), significantly enhances radar resolution.
Additionally, mm-wave FMCW radar offers enhanced privacy in indoor settings, thanks to its high signal attenuation. This technology utilizes subtle chest movements caused by HR and RR to accurately estimate vital signs. The exceptional sensitivity of mm-wave FMCW radar facilitates precise measurements of chest displacement, thereby yielding more accurate estimations of HR and RR.

There are existing datasets of vitals sign recordings by a radar and a reference sensor. Schellenberger, S., et al. \cite{schellenberger2020dataset}, have recorded 24-hour data from 30 healthy subjects by using a 24 GHz continuous wave radar and Electrocardiography (ECG), blood pressure sensor, and Impedance Cardiogram (ICG) as a reference. Five scenarios were carried out, where the aim was to trigger the autonomic nervous system and hemodynamics of the subject. Yoo, Sungwon, et al. \cite{yoo2021radar} have published an FMCW radar recorded dataset of vitals signs for 50 children. Tekleab, Aaron, and Mihai Sanduleanu \cite{tekleab2022vital} have published a dataset for which a mm-wave FMCW radar was used. The test subjects  were 4 children under the age of 13.
To the best of our knowledge, no publicly available dataset utilizes mm-wave FMCW radar for the detection of vital signs in adults.

This paper presents the first comprehensive dataset leveraging mm-wave FMCW radar for the non-invasive monitoring of HR and RR in adults.
Our investigation assesses the radar's accuracy by comparing its measurements against those obtained from the Polar H10, a reference sensor renowned for its precision. 
 Various scenarios, including distance, angle, orientation, and elevated heart rate situations, were considered. Data from ten participants was collected, where four of the participants took part in an elevated heart rate scenario. The experiment aimed to investigate the capabilities of FMCW radar in diverse real-world situations, providing insights into its potential applications. 
The contributions of this dataset are as follows:

\begin{itemize}
    \item \textbf{Diverse Participant Pool:}
    The dataset encompasses recordings from 10 adults, highlighting a variety of demographic groups and physical conditions. This includes:
    \begin{itemize}
        \item Individuals with asthma and an experienced meditator, showcasing the radar's capability to capture data across extreme physiological conditions.
        \item Participants with elevated heart rates due to cardio exercise, further diversifying the dataset's applicability.
    \end{itemize}

    \item \textbf{Diverse Evaluation Scenarios}:

    \begin{itemize}
        \item Emulate real-world applications: Through a series of meticulously designed scenarios—including variations in distance, angle, orientation, and conditions of elevated HR—we aim to emulate real-world applications.
        \item Carefully designed scenarios enable a comprehensive evaluation of radar performance across varying distances, angles, and orientations, reflecting real-world conditions. This approach takes into consideration the specific radar characteristics, features, and antenna patterns, ensuring a thorough assessment of its capabilities in practical applications.
    \end{itemize}

    \item \textbf{Inclusion of Extreme Cases:} Our study uniquely incorporates data from participants with extreme physiological conditions—post-exercise (elevated HR and RR), asthma (high RR), and experienced meditators (low RR)—to demonstrate the radar's effectiveness in real-world scenarios. This approach highlights its capability to accurately detect vital signs across a broad range of conditions.

\item \textbf{Preliminary Validation Through Case Study:} To demonstrate the practical utility and precision of our dataset, we provide initial findings from a detailed case study of one participant. This example highlights the mm-wave FMCW radar's efficacy in accurately monitoring cardiovascular and respiratory systems. These early results serve not only to validate the integrity of the data collected but also to emphasize the radar's potential for non-invasive, precise vital sign monitoring.

\end{itemize}

In the subsequent sections of this paper, we delve into the methodology, participant profiles, scenarios, data formatting, and initial validation efforts in detail. This structure aims to equip other researchers with the necessary information to effectively utilize our dataset, ensuring its reusability and facilitating rigorous testing of its validity for their own studies.

%%%%%%%%%%%%%%%%%%%%%%%%%%%%%%%%%%%%%%%%%%
\section{Methodology}

\begin{figure*}[ht]
    \centering
    \includegraphics[width=12 cm]{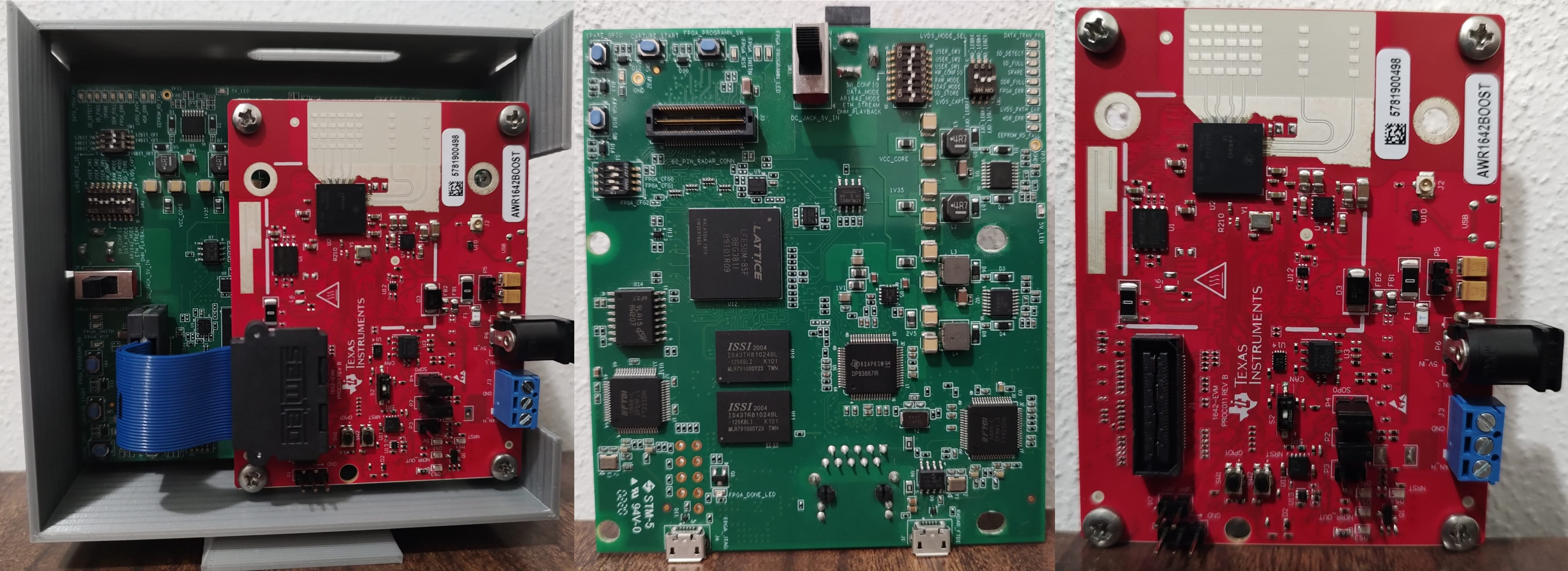}
    \caption{From right to left: AWR1642 EVM- DCA1000- AWR1642 and DCA1000 in the designed radar holder.}
    \label{fig:radar-system}
\end{figure*}
\subsection{Instrumentation and Setting}
In this experiment, we utilized millimeter-wave (mm-wave) Frequency-Modulated Continuous Wave (FMCW) radar. The Texas Instruments (TI) AWR1642BOOST radar automotive radar sensor evaluation module (EVM) was employed for this purpose \cite{TI-AWR1642EVM}. To facilitate the collection of raw data, it was necessary to use the complementary DCA1000 board to capture ADC raw samples of the radar's intermediate frequency (IF) signal\cite{TI-DCA}. Both AWR1642 EVM and DCA1000 used in this experiment can be seen in Fig \ref{fig:radar-system}. Moreover, we designed a radar holder to ease the data collection procedure.
The radar configuration is detailed in Table \ref{table-config}.
This configuration leverages the full potential of the radar system. For instance, by utilizing the entire bandwidth, we aimed to achieve the highest possible range resolution. This principle is mathematically formulated as \(d_{\text{res}} = \frac{C}{2BW}\), where \(C\) denotes the speed of light (\(3 \times 10^8\) m/s), \(d_{\text{res}}\) represents the range resolution, and \(BW\) is the bandwidth utilized by the radar. In our experiment, the full bandwidth of 4 GHz was employed, enhancing the radar's capability to distinguish between closely spaced objects in distance.

It is important to note that the AWR1642 is utilized in a Multiple Input Multiple Output (MIMO) mode. This configuration enables us to harness various MIMO benefits, such as enhanced spectral efficiency and improved signal quality due to diversity gain. These advantages are critical for achieving high-resolution and reliable radar sensing capabilities.

Additionally, as a reference sensor for heart rate measurements, the Polar H10 heart rate sensor was employed. The Polar H10 is regarded as a reliable reference sensor for heart rate measurement, demonstrating superior accuracy over other tested devices and Holter monitors, as confirmed by extensive validation tests.
These tests demonstrated that the H10, when used with the Pro Strap, offered the best heart rate measurement accuracy among all tested systems. Specifically, it detected RR intervals with a precision of within 2 milliseconds accuracy for 92.9\% in running, 99.3\% in cycling, 95.3\% in weight training, and 95.6\% across all activities combined, showcasing its exceptional performance and reliability across various exercise conditions \cite{WP-polar}.

In the data acquisition section, which we elaborate on later, we detail our method for collecting reference values for the respiratory rate. To achieve this, we divided our observation period into one-minute intervals. Participants were instructed to meticulously count their breaths, including both inhalations and exhalations, to ensure precision. Additionally, to validate the values reported by participants, we employed a high-resolution camera specifically focused on the abdominal and chest areas. This approach allowed for a more accurate verification of the respiratory rates provided by the subjects.

\begin{table}  % Change H to htbp for better placement handling
    \centering  % Centers the table
    \newcolumntype{C}{>{\centering\arraybackslash}X}
    \begin{tabularx}{\columnwidth}{CC}  % Adjust width to columnwidth and use two C's if you have two columns
        \toprule
        \textbf{Parameter} & \textbf{Value}\\
        \midrule
        Start Frequency     & 77 GHz\\
        End Frequency       & 81 GHz\\
        ADC Start time      & 6 $\mu$s\\
        ADC Samples         & 250\\
        Sample Rate         & 6250 ksps\\
        Ramp End Time       & 50 $\mu$s\\
        Idle Time           & 7 $\mu$s\\
        % Chirp Count         & 1\\
        Frame Count         & 1200\\
        Frame Periodicity   & 50 ms \\
        Chirp Loop Count    & 128 \\
        RX Gain             & 30 dB\\
        TX Count            & 2\\
        RX Count            & 4\\       
        \bottomrule
    \end{tabularx}
    \caption{Configuration Parameters of AWR1642BOOST.}  % Caption goes before the label
    \label{table-config}
\end{table}

\subsection{Research Experiment Location}

\begin{figure}[htbp]
    \centering
    \includegraphics[width=7cm]{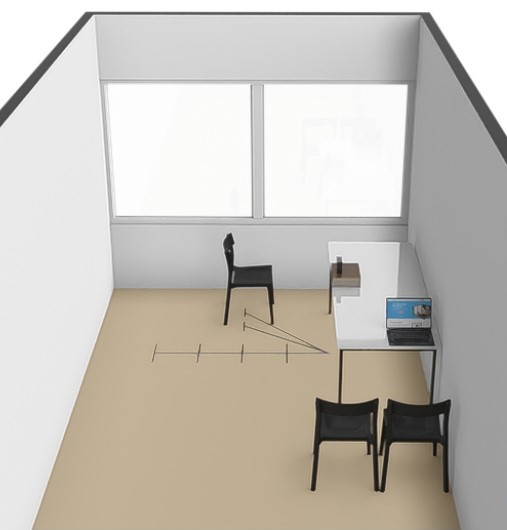}
    \captionsetup{width=7cm} % Set the caption width equal to the image width
    \caption{3D View of the room ZI-5038 and the experimental setup.}
    \label{fig:3d}
\end{figure}

The experiments took place in Room ZI-5038, located within the Pervasive Systems group at the Edge Center on the fifth floor of the Zilverling building, part of the Faculty of Electrical Engineering, Mathematics, and Computer Science (EEMCS) at Twente University, Enschede, the Netherlands. A detailed 3D layout of the room, including the location of the devices used and the positioning of participants, is illustrated in Fig. \ref{fig:3d}.

Minor but important to note, our measurements were conducted during the winter season, a detail of particular importance due to its influence on the clothing thickness of participants. Winter attire, generally bulkier and composed of denser materials, can significantly impact radar signal attenuation and reflection. Such clothing increases signal attenuation, as the materials may absorb or reflect a greater portion of the radar signal, potentially affecting the accuracy of vital sign detection. The dielectric properties of heavier winter clothing could also alter the radar signal's interaction, modifying signal penetration and the characteristics of the reflected signal, including crucial Doppler shifts used for assessing heart rate and respiration. Given these considerations, the outcomes observed in our study, despite the potential for increased signal attenuation due to winter clothing, suggest that the radar-based system possesses robust detection capabilities. Moreover, this implies that experiments conducted with less clothing—resulting in reduced attenuation—might yield even more pronounced results, further demonstrating the efficacy of the mm-wave FMCW radar in diverse conditions.

%%%%%%%%%%%%%%%%%%%%%%%%%%%%%%%%%%%%%%%%%%
\section{Participant Information}

\subsection{Participant Demographics}

%This study received approval from the Computer \& Information Sciences (CIS) Ethics Committee at the University of Twente, with the application number 230671.
A total of 10 participants were recruited for the measurement sessions, comprising an equal distribution of 5 males and 5 females. Prior to their involvement, all participants were fully briefed on the procedures and objectives of the study and provided informed consent by signing a consent form.

\begin{figure}[t]
    \centering
    \includegraphics[width=8cm]{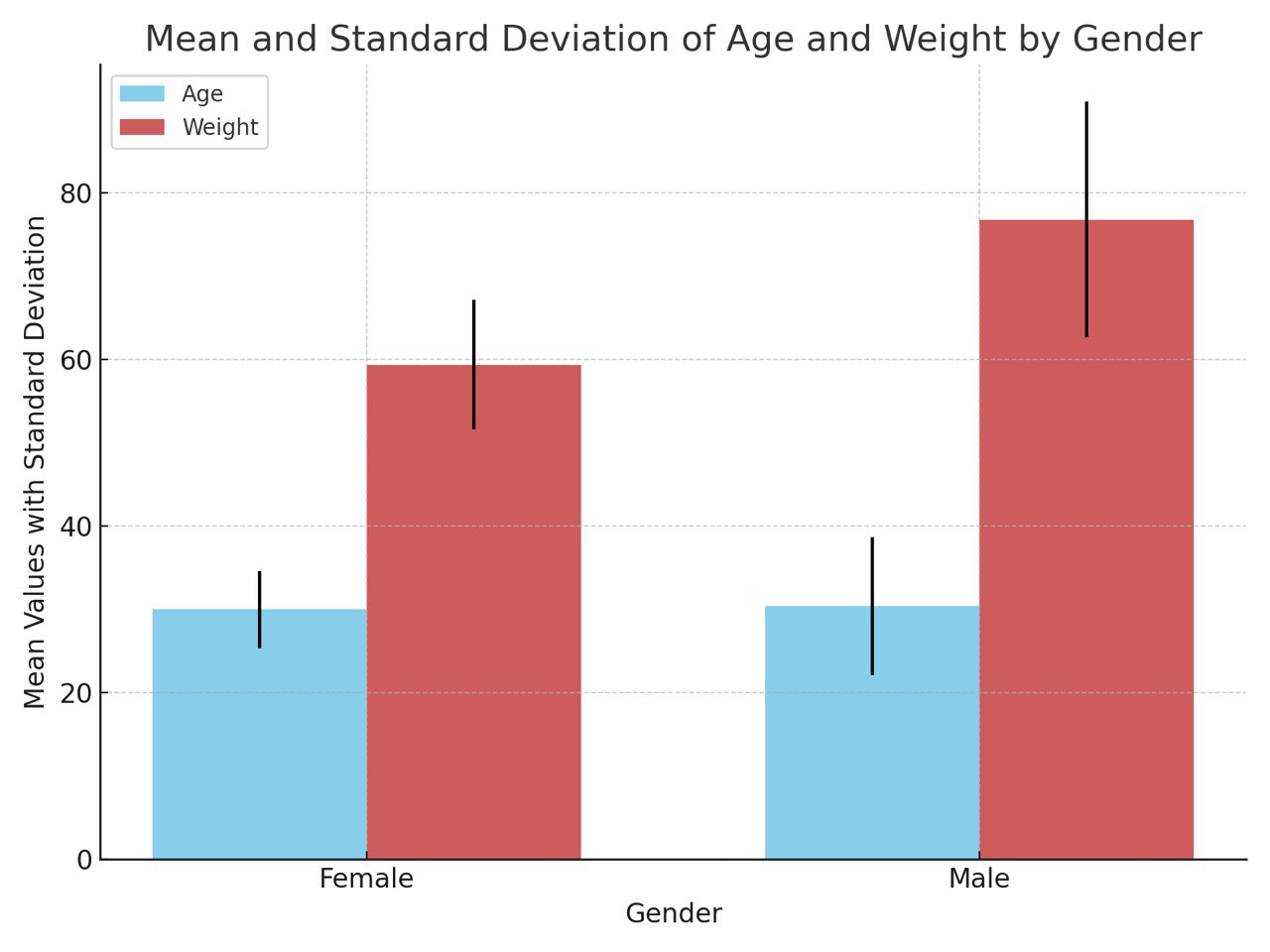}
    \captionsetup{width=8cm}  % Set the caption width equal to the image width
    \caption{Mean and Standard deviation for age and weight among participants.}
    \label{fig3}
\end{figure}

The participants' weights and ages were recorded, with the mean weight being $\bar{x}_{\text{weight}}= 68.1 \text{ kg}$ and its standard deviation $\sigma_{\text{weight}} = 14.14 \text{ kg}$. Similarly, the mean age was $\bar{x}_{\text{age}} = 30.2 \text{ years}$, with a standard deviation of $\sigma_{\text{age}} = 6.32 \text{ years} $. The distribution of both age and weight is summarized in Fig. \ref{fig3} for a comprehensive overview.
Each of the 10 participants took part in the distance, orientation, and angle scenarios to assess the radar's performance under varying conditions. Furthermore, a subset of four participants also engaged in the elevated HR scenario, details of which are tabulated in Table \ref{table1}. This diverse participant pool ensures a broad representation of data for analyzing the radar's effectiveness across different demographic groups and physical conditions.

\begin{table*}

\newcolumntype{C}{>{\centering\arraybackslash}X}
\begin{tabularx}{\textwidth}{CCCCC}
\toprule
\textbf{Participant number}	& \textbf{Gender}& \textbf{Age}&  \textbf{Weight} &\textbf{Engaged Scenarios}\\
\midrule
%Entry 1		& Data		&	& Data\\
%Entry 2		& Data		&	& Data \textsuperscript{1}\\
Participant 1& Male& 22& 83&Distance, Orientation, Angle \\\\
Participant 2$^{\dagger}$& Male & 27 & 70 &Distance, Orientation, Angle, Elevated \\\\
Participant 3& Female& 30& 60 & Distance, Orientation, Angle, Elevated\\\\
Participant 4& Female & 25 & 63 & Distance, Orientation, Angle, Elevated\\\\
Participant 5$^{*}$& Female& 26 & 54&Distance, Orientation, Angle \\\\
Participant 6\textsuperscript{*}& Male&  31&98.5 &Distance, Orientation, Angle, Elevated\\\\
Participant 7& Female&33 &70 &Distance, Orientation, Angle \\\\
Participant 8& Male &28 & 63& Distance, Orientation, Angle\\\\
Participant 9& Female&36 & 50& Distance, Orientation, Angle\\\\
Participant 10& Male& 44&69.5 & Distance, Orientation, Angle\\\\
\bottomrule
\end{tabularx}
\noindent{\footnotesize{\textsuperscript{$^{\dagger}$} Participant with extensive experience in meditation.}}\\
\noindent{\footnotesize{\textsuperscript{$^{*}$} Participant diagnosed with asthma.}}
\caption{List of participants as well as their gender, age and engagement in different scenarios.}
\label{table1}
\end{table*}

\subsection{Special Participant Profiles}

In our study, we specifically focused on testing the radar's performance across a range of physiological extremes to ensure its utility in diverse real-world situations. We selected participants who represent the extremes of the RR spectrum: individuals with asthma, known for higher than average RRs, particularly during acute attacks, and a participant experienced in meditation, characterized by significantly lower RRs.

The inclusion of asthmatic participants allowed us to test the radar's sensitivity to faster breathing patterns, as they tend to have a higher baseline RR compared to non-asthmatics, a factor well-documented in existing research \cite{asthma}. On the other hand, our meditative participant, with RR as low as 3.5 breaths per minute (BPM), provided a unique case for examining the effects of deep relaxation and parasympathetic activation on respiratory function. Studies, such as those by Sony et al. (2019), have shown that Breath Rate Variability (BRV) is a promising metric for distinguishing meditators from non-meditators \cite{Med-participant}.
By incorporating these cases, we aimed to validate the radar's ability to adapt to and accurately detect vital signs in a varied population, affirming its potential for broad clinical application.

%%%%%%%%%%%%%%%%%%%%%%%%%%%%%%%%%%%%%%%%%%
\section{Data Collection Scenarios}

The primary objective of these measurements is to explore the capabilities of FMCW radar in non-contact health monitoring, with the ultimate goal of developing a standalone product capable of autonomous health monitoring.
In practical applications, FMCW radar has the potential for a wide array of uses, including vital sign detection and activity recognition. To fully realize this potential, it's crucial to simulate realistic conditions that the radar might encounter in real-world scenarios. This includes situations where the radar is not ideally positioned relative to the subject, such as not facing the chest area directly.

To address this, we have devised a series of distinct scenarios to rigorously test the radar's accuracy and reliability under various conditions. These scenarios involve varying the distance, angle, and orientation of human participants relative to the radar.
Additionally, our study extends to assessing the radar's performance under specialized conditions. This includes gathering data from individuals with asthma to understand the radar's sensitivity to varying respiratory patterns. We also focus on scenarios involving elevated heart and breathing rates, achieved by having participants engage in physical activities like climbing stairs. These scenarios are vital for evaluating the radar's effectiveness in dynamic, real-world conditions where physiological parameters may deviate from the norm due to various activities or health conditions.

Such a comprehensive evaluation aims to deepen our understanding of the radar's performance across different settings and identify strategies to maintain or enhance accuracy in less-than-ideal conditions.
Below, we detail the purpose and methodology of each scenario, providing insights into our experimental approach and the rationale behind it.

\subsection{Distance Scenario}

The mmWave FMCW radar operates in a frequency range that uniquely balances advantages against challenges. A significant advantage of mmWave frequencies is the enhancement of privacy, given that signals typically do not penetrate walls, thus minimizing the risk of data interception. Nonetheless, this characteristic contributes to higher signal attenuation as the distance from the radar source increases, presenting a unique set of challenges \cite{DataScenario-1}.

As distance grows, the power of the desired signal tends to wane, and concurrently, the noise level within the signal may ascend. This dynamic shift leads to a reduced Signal-to-Noise Ratio (SNR), complicating the task for signal processing algorithms to accurately estimate vital signs such as HR and RR. Recognizing the critical nature of these challenges, this scenario is dedicated to scrutinizing the radar's efficacy and the precision of HR and RR estimations across varying distances. Grasping the extent to which signal attenuation influences data quality is paramount for the development of robust health monitoring applications leveraging FMCW radar technology.

In this experimental setup, participants were seated at specific distances from the radar — 40 cm, 80 cm, 120 cm, and 160 cm — with their chest area oriented directly towards the radar apparatus. To ensure the purity of data collection, participants were advised to remain motionless, thereby reducing the potential for data interference.

Data was meticulously recorded in four separate one-minute segments for each participant at every prescribed distance. This methodical approach facilitated a comprehensive assessment of how distance impacts both the quality of the signal and the accuracy of the vital sign estimations derived from it. This endeavor aims not only to illuminate the challenges posed by increasing distances but also to identify potential strategies to mitigate these effects, ensuring the radar's applicability in a wide array of health monitoring contexts

\subsection{Orientation Scenario}
In practical applications, individuals may assume various orientations relative to the radar. Understanding the radar's capabilities and limitations across all possible orientations is crucial, particularly for deployment in healthcare settings, vehicles, or smart home systems where users might not directly face the radar. To address this, we investigate the impact of orientation on radar performance while maintaining a consistent distance.
Participants were seated at a fixed distance of 80 cm from the radar, facing in different directions rather than directly towards it. This setup allows us to systematically collect data as the front, back, left side, and right side of the participant's body are oriented towards the radar. For each orientation, we record four separate one-minute data sessions.

The orientation of the participant relative to the radar is significant for several reasons. The Radar Cross Section (RCS) varies significantly with body orientation. Typically, the front of the body presents a larger RCS due to the greater surface area facing the radar, resulting in stronger reflected signals. Conversely, the sides and back exhibit a smaller RCS, potentially leading to weaker or differently characterized reflections.

Different orientations also affect signal penetration, reflection, and absorption, altering the characteristics of the received signal. The Line of Sight (LOS) signal carries different information depending on the orientation, and multipath propagation dynamics change accordingly. Orientations other than directly facing the radar are expected to significantly reduce the SNR of the chest displacement pattern, impacting the accuracy of vital sign detection. Additionally, different orientations influence the object's position relative to the main lobe of the antenna pattern and affect Doppler shifts due to chest movements. By understanding these dynamics, we can optimize radar performance and develop algorithms that are robust across various orientations, enhancing the radar's applicability in diverse settings.

\begin{figure}[t]
    \centering
    \includegraphics[width=8.9cm]{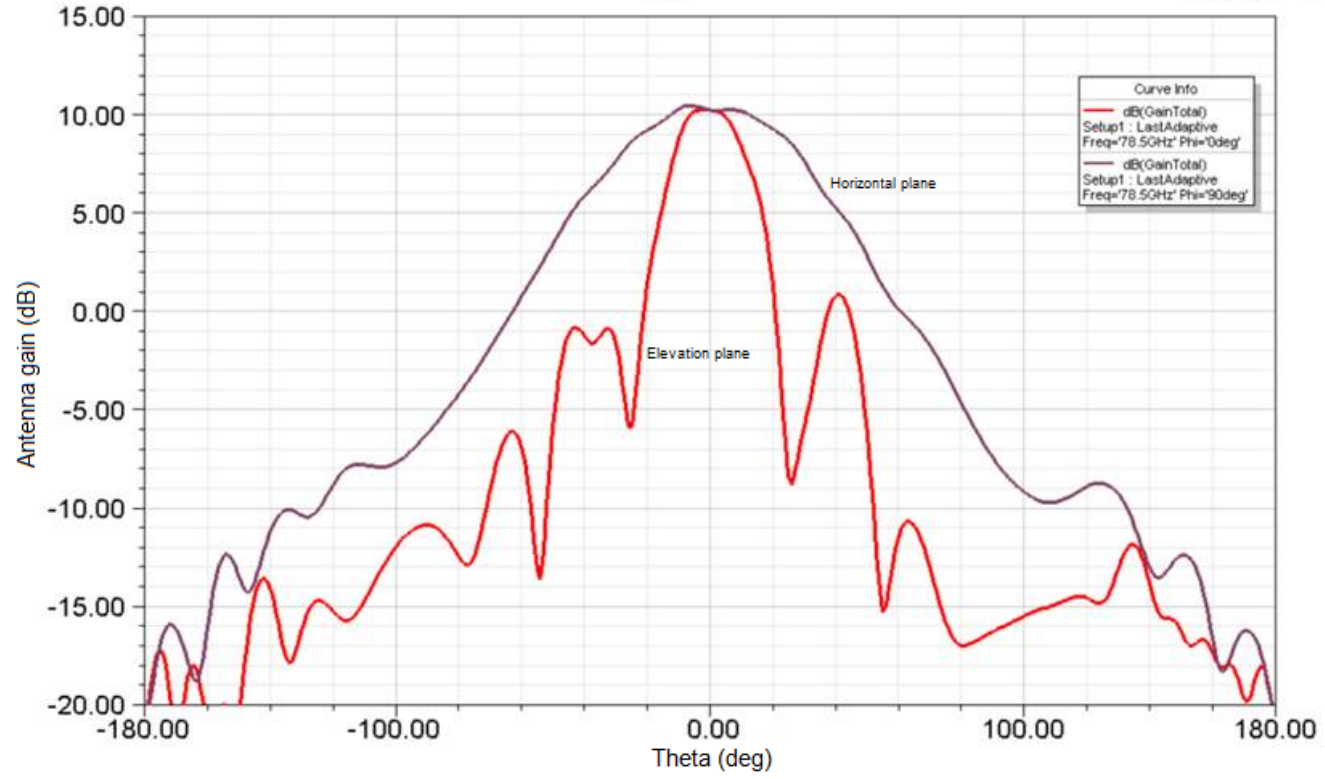}
    \captionsetup{width=8.9cm}  % Set the caption width equal to the image width
    \caption{Antenna pattern of AWR1642 EVM \cite{TI-AWR1642EVM}}
    \label{fig1}
\end{figure}

\subsection{Angle Scenario}

In real-world applications, it's unlikely that subjects will always be perfectly aligned with the radar's direct line of sight. To mimic these realistic conditions and assess the radar's adaptability, testing at various angles is essential.

The antenna pattern of the AWR1642 EVM, as depicted in Fig. \ref{fig1}, illustrates how antenna gain varies with the angle (theta) \cite{TI-AWR1642EVM}. This pattern indicates that antenna gain is maximized at 0 degrees and diminishes as the angle widens. Consequently, the most precise vital sign estimations are anticipated at 0 degrees. As the angle deviates to 30 and 45 degrees, a reduction in signal strength is expected, following the antenna gain pattern. This reduction could lead to a lower SNR of the received signal, potentially impacting the accuracy of HR and RR estimations.

To investigate these effects, data were collected from participants seated at a consistent distance of 80 cm from the desk, on which radar was placed. Participants were positioned with 0, 30, and 45 degrees of deviation on the horizontal plane relative to the radar's normal axis. For each angular setting, four separate one-minute measurements were captured using the radar.
It's important to note that participants were oriented toward the desk, not directly facing the radar. This setup more closely simulates real-life scenarios where individuals may not always be ideally positioned toward the radar. Such an approach not only aids in evaluating the radar's performance under varied angular conditions but also contributes to understanding how radar signals reflect off different surfaces and body orientations. Enhancing our understanding of these dynamics is crucial for improving the robustness of algorithms designed for vital sign detection in less-than-ideal conditions.

\subsection{Elevated HR Scenario}

The elevated HR and RR scenario is designed to assess the radar's capability under conditions of increased physiological activity, such as physical exertion or stress. This evaluation is crucial for ensuring the radar's applicability in real-world scenarios beyond calm, resting states. It particularly focuses on the radar's sensitivity and accuracy in tracking rapid changes in heart rate and respiration rate, which are vital for applications requiring immediate health monitoring or stress detection.

In this scenario, four participants, including one individual with asthma, were selected to induce elevated heart and respiration rates through physical activity. They were instructed to quickly ascend the stairs to the fifth floor of the Zilverling building within the EEMCS faculty at Twente University. Immediately after reaching the designated floor, participants positioned themselves 80 cm in front of the radar, ensuring they were facing towards it, to commence data collection. Two separate one-minute data segments were recorded in this state to capture the elevated physiological rates.
To further assess the radar's performance as heart and respiration rates began to stabilize, participants were not given a rest period. Instead, after the initial two minutes of recording—during which their heart and respiration rates naturally started to return towards normal levels—they were immediately asked to undertake the stair-climbing activity once more. This repetition aimed to induce a second increase in heart and respiration rates, after which another two separate one-minute data recordings were captured using the radar. This iterative approach allows for a nuanced understanding of the radar system's adaptability to rapid physiological fluctuations, offering insights into its potential reliability and effectiveness in dynamic, real-life applications.

%%%%%%%%%%%%%%%%%%%%%%%%%%%%%%%%%%%%%%%%%%

\begin{figure}[t]
    \centering
    \includegraphics[width=8.8cm]{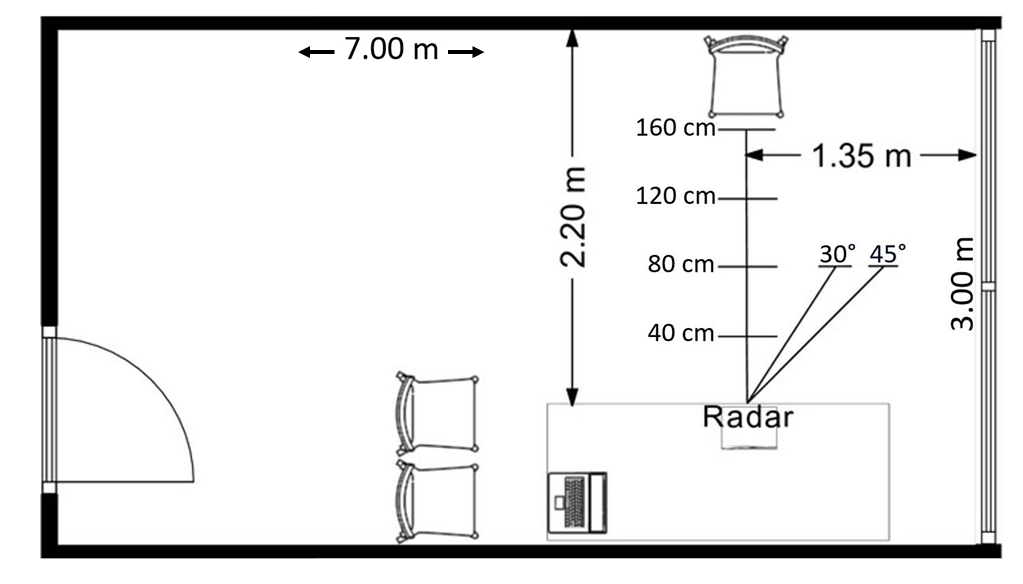}
    \captionsetup{width=8.8cm}  % Set the caption width equal to the image width
    \caption{The floor map: dimensions and the experimental setup.}
    \label{floormap}
\end{figure}

\section{Experimental Setup}

\begin{figure}[t]
    \centering
    \includegraphics[width=8.6cm]{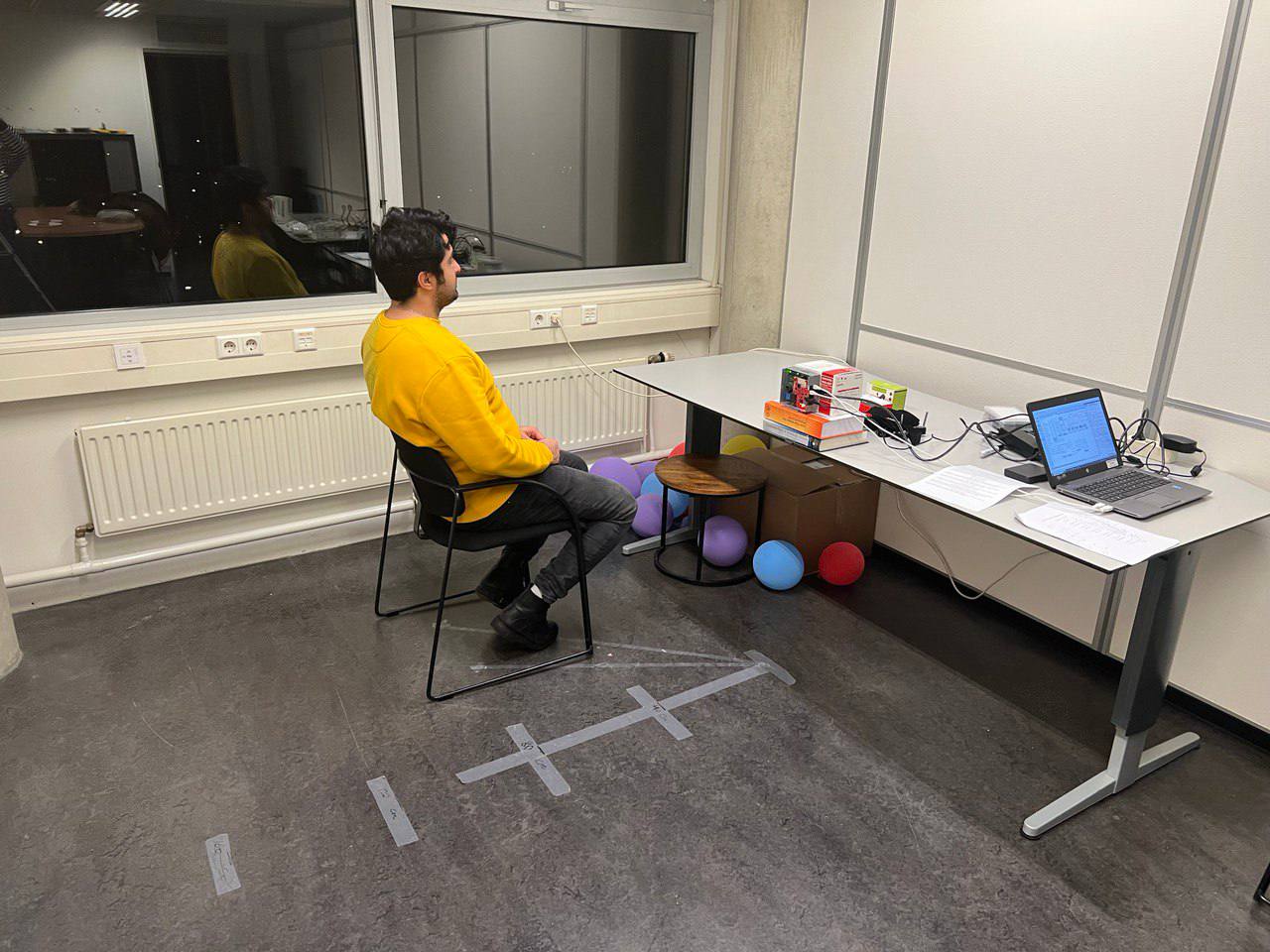}
    \captionsetup{width=8.6cm}  % Set the caption width equal to the image width
    \caption{Participant Demonstrating the Angle Scenario within the Experimental Setup.}
    \label{participant}
\end{figure}

The radar system, comprising the AWR1642 EVM, DCA1000, and radar holder, was positioned on an office table. Elevation adjustments were implemented to align the radar antenna array with the chest of the seated individual, ensuring consistency across participants. The radar system's location remained constant throughout the study. The participants were sitting on a chair, approximately 45 centimeters from the ground, and the radar was placed in a way that the antennas were located approximately 90 centimeters from the ground. A laptop computer, placed on the same table, established connections with the radar system according to the guidelines outlined in the tutorial \cite{TI-DCA}. To facilitate smooth transitions between scenarios, demarcations were made on the floor using duct tape to indicate distances of 40 cm, 80 cm, 120 cm, and 160 cm (distance scenario), along with angles of 30 and 45 degrees at a distance of 80 cm from the radar (angle scenario). The setup remained consistent for the elevated scenario, ensuring uniformity across measurements. For the orientation scenario, the chair was strategically rotated to adjust the participant's position relative to the radar, accommodating the varied orientation required for this specific set of experiments. To provide a clear understanding of the spatial arrangement and distances utilized in each experimental scenario, a detailed floor map of the room and the entire setup is included. This visual representation can be found in Fig. \ref{floormap}, offering an invaluable perspective on the physical context of our measurements.
Chair movements to specified distances were guided by the tape, ensuring precise positioning. Upon seating, distance validation included a visual alignment check between the tape indication and the participant's chest. In Fig \ref{participant}, a participant can be seen while collecting data for the angle scenario.

\section{Data Acquisition}

\subsection{Data recording procedure}

Adjacent to the participant's right side, three individuals, out of the radar's line of sight, facilitated the experiment. One operated the laptop for radar measurements through the TI mm-wave Studio software.
The TI mmWave Studio serves as the software for radar and DCA1000 configuration, as well as data recording. Additional software prerequisites included Matlab Runtime Engine V8.5.1 or higher and Code Composer Studio V7.1 or higher. Instructions on the connections and software utilization can be found in the DCA1000 training tutorial \cite{TI-DCA}.
The second ensured synchronized measurements via the Polar Beat application (for more information check the Polar website).
The Polar Beat software application on an Android phone facilitated the recording of all measurements. The heart rate sensor was paired with the Polar Beat app through Bluetooth to ensure accurate and synchronized heart rate data collection.
Finally,  the third counted breaths and recorded 1-minute high-resolution video recordings of the participant's chest area. Video documentation was initiated in case breath counts required verification. 

At the end of each 1-minute measurement, the participant reported the breath count, and it was cross-validated with the third person's count. Later verification through recorded high-resolution videos was done to ensure the accuracy of the reported breath count.

\subsection{Participant Preparation}

Prior to commencing measurements, participants were provided with an informed consent form, which, upon signing, was followed by a concise explanation of the primary objectives of the study. Subsequently, participants were directed to affix the Polar H10 heart rate sensor directly onto their skin, positioned just below the sternum. Verification of a secure connection between the heart rate sensor and the phone running the Polar Beat application was conducted to ensure reliable heart rate measurements. Following the sensor setup, participants were guided to take a seat on a chair positioned in front of the radar, corresponding to the first measurement scenario. Participants were instructed to minimize body movements during measurements, with no specific guidelines regarding breathing rate—participants were neither directed to maintain a normal nor abnormal rate. However, they were prompted to count their breaths during the 1-minute measurements and articulate the results aloud upon completion for recording purposes. Note that the participants were asked to account for both exhalation and inhalation, ensuring the inclusion of incomplete breaths in our analysis. Post each scenario, participants were asked to stand while the chair was relocated to the next setup for subsequent measurements. This procedural approach maintained consistency across scenarios and facilitated a smooth transition between measurement configurations. It should be noted that all distances mentioned are measured from the chest area to the antenna patch, ensuring precision in the spatial configuration of our experiments.

\subsection{Data format}

\subsubsection{Mm-wave FMCW Radar}

\begin{table*} 

\newcolumntype{C}{>{\centering\arraybackslash}X}
\begin{tabularx}{\textwidth}{CCCCC}
\toprule
\textbf{Folder/Subfolder Hierarchy}	& \textbf{Sub folders}& \textbf{Recordings folder}&  \textbf{File name} &\textbf{Format}\\
\midrule
%Entry 1		& Data		&	& Data\\
%Entry 2		& Data		&	& Data \textsuperscript{1}\\
BR\_Ref\_Values&---& ---&BR\_Ref\_Values&CSV \\\\
\hline
HR\_Ref\_Values/ & Participant X/& & & \\\\
 .../1. Distance Scenario/ & 40 cm, 80 cm, 120 cm, 160 cm& 1,2,3,4&  R1, R2, R3, R4& CSV\\\\
 
 .../2. Orientation Scenario/ & 80 cm- back, 80 cm- Front side, 80 cm- left side, 80 cm- Right side& 1,2,3,4&  R1, R2, R3, R4& CSV\\\\
 
 .../3. Angle Scenario/ & 0 deg- 80 cm, 30 deg- 80 cm, 45 deg- 80 cm & 1,2,3,4&  R1, R2, R3, R4& CSV\\\\
 
 .../4. Elevated/$^{*}$ & ---  & --- &  R1, R2, R3, R4& CSV\\\\
\hline
Radar data/ & Participant X/& & & \\\\
 .../1. Distance Scenario/ & 40 cm, 80 cm, 120 cm, 160 cm& 1,2,3,4&  data\_Raw\_0& Binary (bin)\\\\
 
 .../2. Orientation Scenario/ & 80 cm- back, 80 cm- Front side, 80 cm- left side, 80 cm- Right side& 1,2,3,4&   data\_Raw\_0& Binary (bin)\\\\
 
 .../3. Angle Scenario/ & 0 deg- 80 cm, 30 deg- 80 cm, 45 deg- 80 cm & 1,2,3,4&   data\_Raw\_0& Binary (bin)\\\\
 
 .../4. Elevated/$^{*}$ & ---  & 1,2,3,4 &  data\_Raw\_0 & Binary (bin)\\\\
\hline

\bottomrule
\end{tabularx}
\noindent{\footnotesize{\textsuperscript{*} Only for participants 2, 3, 4, and 6.}}
\caption{Simplified Dataset Structure}
\label{Structure}
\end{table*}

\begin{figure}[thb]
\hspace{-0.5cm}
    \centering
    \includegraphics[width=9.1cm]{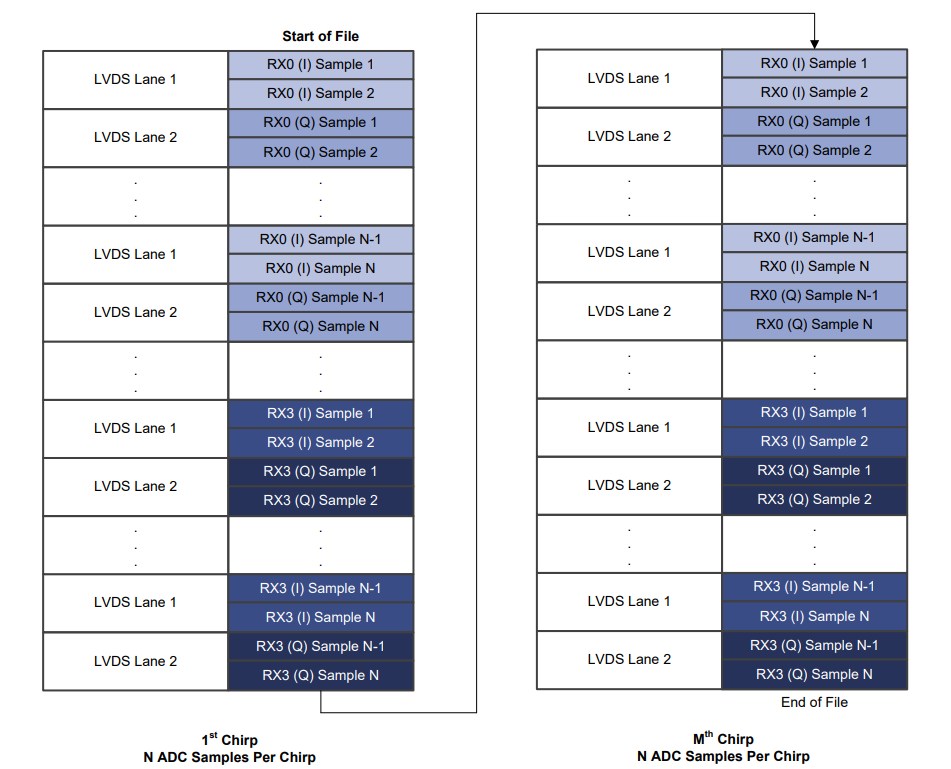}
    \captionsetup{width=9cm}  % Set the caption width equal to the image width
    \caption{xWR16xx/IWR6843 Complex Data Format Using DCA1000 \cite{TI-DCA}.}
    \label{fig2}
\end{figure}

In the data collection process using the FMCW Radar, the file size is determined by the number of ADC samples ($N_{ADC}$), the number of receive channels ($ N_{RX}$), the number of frames ($N_{F}$), the number of chirps ($N_{C}$), and the number of bytes per sample ($B_{S}$). The formula for calculating the expected file size, reflecting the radar configuration, is presented below:
% \begin{equation}
%    \text{Total Size in Bytes} = N_{ADC} \times N_{RX} \times N_{Frames} \times N_{Chirps} \times B_{Sample}
% \end{equation}
\begin{equation}
   \text{Total Size} = N_{ADC} \times N_{RX} \times N_{F} \times N_{C} \times B_{S},
\end{equation}

where the resulting number represents the total file size in bytes. This formulation succinctly captures the total approximate memory required to store the captured ADC data.
For our specific setup, with 250 ADC samples, 4 receive channels, 1200 frames, 128 chirps, and 4 bytes per sample (to account for IQ demodulation, where ``I'' stands for in-phase and ``Q'' for quadrature, each contributing 2 bytes per sample), the calculated file size is approximately 614,400,000 bytes, or 585.94 megabytes. The slight discrepancy observed, with the actual file size being around 600 megabytes, is within acceptable limits. This difference can often be attributed to file system overhead, the inclusion of metadata within the file, or the formatting of the data storage. Such a marginal discrepancy is not uncommon and does not generally indicate any issues with the data integrity or the radar's performance. It is important to note that this calculation assumes a seamless data collection process without any additional data or headers that might be included in the file.
The order of IQ samples from each receiver and LVDS lanes are explained in detail in Fig \ref{fig2}.
More details can be found in the DCA1000EVM user's guide and TI mmWave radar application report \cite{TI-DCA, TI-AWR1642EVM}.

\begin{figure*}
\hspace{-0.5cm}
    \centering
    \includegraphics[width=17 cm]{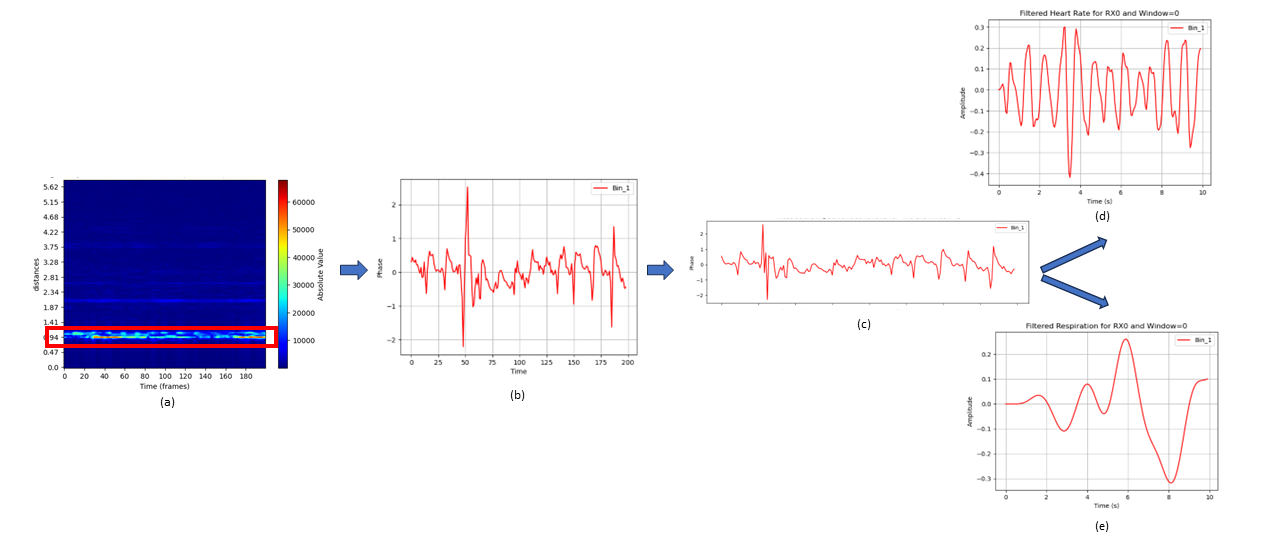}
    \caption{Signal processing chain- (a) Range map, (b) Unwrapped phase Signal, (c) Chest displacement signal, (d) Filtered HR signal, (e) Filtered RR signal for the first Window (200 frames= 10 seconds).}
    \label{SP-chain}
\end{figure*}

\subsubsection{Polar H10}
The reference heart rate measurements were obtained using the Polar Beat application, with synchronization facilitated by the Polar Flow application. Saving and accessing the recordings were accomplished through the Polar Flow website. Within the website's diary section, data could be preserved by utilizing the "Export session" option and selecting the export training session as a CSV file. The approximate size of the file is 2 KB. 
The CSV file structure comprises three initial rows containing recording details. Row 1 delineates parameter names, including name, sport, date, start time, duration, etc. Row 2 corresponds to parameter values aligned with the categories in Row 1. Row 3 specifies parameter names such as sample rate, time, HR (bmp), etc. Subsequent rows consist of parameter values related to the parameters outlined in Row 3, with only two parameters - time and HR in beats per minute (bmp). The format of rows 4 to 63+ is \textit{",00:00:01,73,,,,,,,,,"}, where the time (\textit{00:00:01}) of the measurement is in column 2, and the heart rate (\textit{73}) is recorded in column 3. Each recording was made for the time between \textit{00:00:00}, and stopped between time stamps \textit{00:00:59} and \textit{00:01:02}, hence the amount of rows in the recorded CSV files vary between 63 and 66 rows.

Ultimately, following verification with recordings from a high-resolution camera, the breath count reported for each participant was meticulously documented in an Excel sheet. To facilitate understanding of the dataset's structure, a simplified version is illustrated in Table \ref{Structure}, showcasing how our bin and CSV files are organized.

\subsection{Data Availability}

This study received approval from the Computer \& Information Sciences (CIS) Ethics Committee at the University of Twente, with the application number 230671.
The dataset is available at the 4TU.ResearchData under the CC BY-NC-SA 4.0 license with the DOI: https://doi.org/10.4121/48acba04-96bc-4131-b52f-9e18458ad92b \cite{dataset}.

\section{Dataset Validation and Preliminary Findings}

Our study leverages the mmWave FMCW radar's capability to emit and receive signals, culminating in the generation of an Intermediate Frequency (IF) signal through signal mixing. This process results in ADC samples of each transmitted chirp represented as complex numbers due to IQ demodulation, laying the groundwork for our signal processing.

To validate our dataset, this section provides a concise overview of the signal processing steps applied to a specific example: participant 1 in the distance scenario (80 cm), during the second recording. A general outline of the procedures to estimate HR and RR is illustrated in Fig \ref{SP-chain}.
In this case study, we observed a 50-second timeframe, employing a window size (W) of 200 frames (equivalent to 10 seconds) and a sliding window (SW) of 20 frames (1 second). It's noteworthy that increasing both the window size and the sliding window duration could potentially enhance the results and accuracy. However, these parameters were selected to optimize the estimation of the radar's performance, particularly with an eye toward real-time monitoring capabilities.
The HR/ RR extraction approach employed in this signal processing example relies on simple band-pass filtering (BPF) and Fast Fourier Transform (FFT) techniques. While it's recognized that more sophisticated signal processing methods could yield higher accuracy and improved performance, the focus of this article is not on the depth of signal processing intricacies. Future work will explore how advanced processing techniques can achieve greater precision. Below, we provide a summary of the signal processing steps used to estimate HR and RR for Participant 1, within the distance scenario and recording number 2, given $W=200$ and $SW=20$.

With each frame comprising 128 chirps and each chirp yielding 250 ADC samples, our methodology first entails organizing these samples across all four receivers. This structured approach forms a data cube incorporating data from 4 antennas, 250 samples for each chirp, and a total of 1200 frames multiplied by 128 chirps. Following this organization phase, the FFT is applied to each column—representing the 250 samples of each chirp. This step is crucial for generating a range map \ref{SP-chain} (a).
Within the range map, higher absolute values within range bins in each chirp indicate the presence of an object or subject within the scene. Identifying the maximum range bin allows us to concentrate on the signal from the target (participants), facilitating the unwrapping of the phase to ascertain the chest displacement signal.

Subsequent signal processing steps include the elimination of static objects, offsets, clutters, and nonlinearities in the radar transceiver, leading to the extraction of the IF signal's phase Fig \ref{SP-chain} (b). Through careful filtering, noise and impulsive movements are removed, revealing the chest displacement pattern Fig \ref{SP-chain} (c). The chest displacement signal then goes through bandpass filtering to isolate breathing and heart rate frequencies, yielding a time-domain signal including only HR Fig \ref{SP-chain} (d) or RR Fig \ref{SP-chain} (e) signal.

For the purpose of this study, which prioritizes data collection and description over in-depth signal processing analysis, a simple FFT method is employed for preliminary validation. This approach yields heart rate and respiration rate estimates, with Mean Absolute Percentage Error (MAPE) values of 4.44 \% for RR and 7.36 \% for HR in this case study considering observation time of 50 seconds, W=10 sec, and SW=1 sec.

\begin{figure}[t]
\hspace{-0.5cm}
    \centering
    \includegraphics[width=8.9cm]{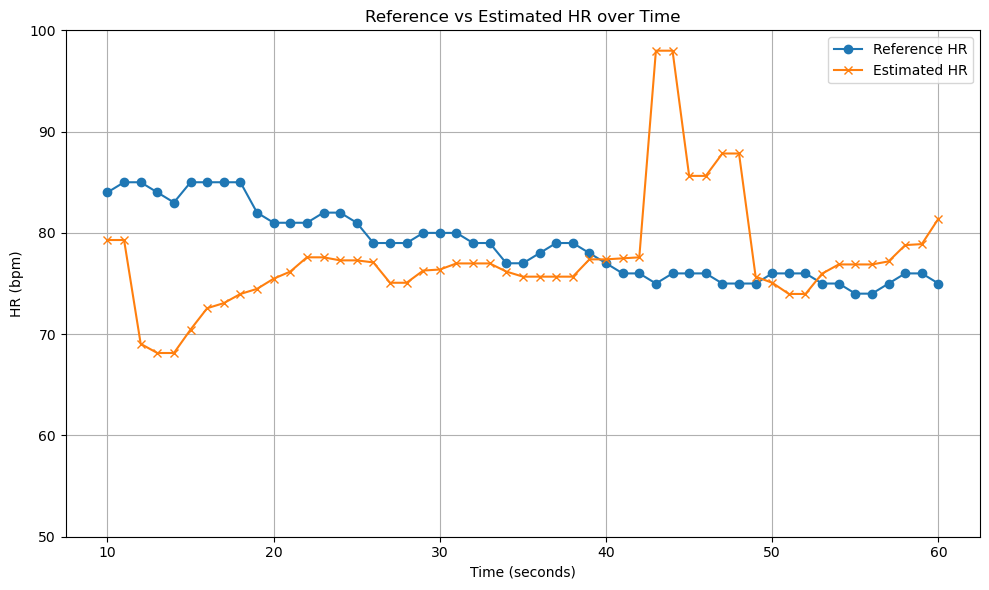}
    \captionsetup{width=8.9cm}  % Set the caption width equal to the image width
    \caption{Comparing estimated values and reference values for HR in 50 seconds observation time}
    \label{HR-Ref}
\end{figure}

To assess the suitability of this method for real-time HR monitoring, Fig \ref{HR-Ref} displays a comparison between HR values estimated using mm-wave FMCW radar and those obtained from the reference Polar belt over a 50-second observation period. In this analysis, a sliding window of 1 second was employed to match the update frequency of the Polar belt, facilitating a closer approximation to real-time monitoring. It's evident that utilizing longer windows and adjusting the sliding window parameters could further improve the accuracy of the estimated HR values. While the graph does present some outliers, these discrepancies highlight opportunities for refinement through more sophisticated signal processing techniques, potentially enhancing the precision of HR estimations significantly.

\subsection{Conclusion}
In this pioneering study, we introduce a comprehensive dataset derived from mm-Wave FMCW radar, adeptly capturing vital signs across diverse physiological states, including scenarios with extreme heart and respiratory rates. Through rigorous validation, we've ensured the dataset's accuracy and reliability, highlighting its potential as a foundational resource for the field. While directly fostering advancements in non-invasive monitoring technologies, this dataset offers invaluable data for researchers aiming to develop and refine health monitoring solutions, contributing to the evolution of healthcare methodologies and practices.

% \funding{This research received no external funding'' or ``This research was funded by NAME OF FUNDER grant number XXX.'' and  and ``The APC was funded by XXX''. Check carefully that the details given are accurate and use the standard spelling of funding agency names at \url{https://search.crossref.org/funding}, any errors may affect your future funding.}

% \institutionalreview{This study received approval from the Computer \& Information Sciences (CIS) Ethics Committee at the University of Twente, with the application number 230671.}

% \informedconsent{Informed consent was obtained from all subjects involved in the study.}

% \dataavailability{The dataset is available at the 4TU.ResearchData under the CC
% BY-NC-SA license with the DOI: } 

% \conflictsofinterest{The authors declare no conflicts of interest.} 

\bibliographystyle{IEEEtran}
\bibliography{main.bib}  % Here the bibliography (references.bib) file name

\end{document}